\documentclass[prl,superscriptaddress,twocolumn,showpacs,preprintnumbers,amsmath,amssymb]{revtex4}

\usepackage{epsfig}
\usepackage{dcolumn}
\usepackage{bm}

\def\simge{\mathrel{%
   \rlap{\raise 0.511ex \hbox{$>$}}{\lower 0.511ex \hbox{$\sim$}}}}
\def\simle{\mathrel{
   \rlap{\raise 0.511ex \hbox{$<$}}{\lower 0.511ex \hbox{$\sim$}}}}

\newcommand{\muplus}{\mu^+}
\newcommand{\muminus}{\mu^-}
\newcommand{\piplus}{\pi^+}
\newcommand{\piminus}{\pi^-}
\newcommand{\kplus}{K^+}

\newcommand{\jpsi}{J/\psi}

\newcommand{\Vzero}{V^0}
\newcommand{\kshort}{K_S^0}
\newcommand{\lam}{\Lambda^0}
\newcommand{\kstarzero}{K^{*0}}
\newcommand{\kstarplus}{K^{*+}}
\newcommand{\bzero}{B^0}
\newcommand{\bplus}{B^{+}}
\newcommand{\lambdab}{\Lambda_{b}^{0}}

\newcommand{\jpsimumu}{\jpsi \rightarrow \muplus \muminus}

\newcommand{\kspipi}{\kshort \rightarrow \piplus \piminus}
\newcommand{\lmppi}{\lam \rightarrow  p \piminus}
\newcommand{\kstarzerokpi}{\kstarzero \rightarrow \kplus \piminus}
\newcommand{\kstarpluskspi}{\kstarplus \rightarrow \kshort \piplus}

\newcommand{\bzerojpsiks}{\bzero \rightarrow \jpsi \kshort}

\newcommand{\lambdabjpsilm}{\lambdab \rightarrow \jpsi \lam}

\newcommand{\PDL}{ct}
\newcommand{\ctauData}{ct_i}
\newcommand{\ctauErrData}{\sigma^{ct}_i}
\newcommand{\massData}{m_i}
\newcommand{\massErrData}{\sigma^m_i}
\newcommand{\ctau}{c\tau}

\newcommand{\LxyFromJpsi}{L_{xy}^{\Vzero}}
\newcommand{\LxyErrFromJpsi}{\sigma_{\LxyFromJpsi}}
\newcommand{\LxySigFromJpsi}{\LxyFromJpsi/\LxyErrFromJpsi}
\newcommand{\LxyB}{L_{xy}^b}
\newcommand{\chisq}{\chi^2}
\newcommand{\pt}{p_{\rm T}}
\newcommand{\ptB}{\pt^{b}}

\newcommand{\mum}{\mu {\rm m}}

\newcommand {\GeVc} {~{\rm GeV}/c}

\newcommand {\GeV} {~{\rm GeV}/c^2}
\newcommand {\MeV} {~{\rm MeV}/c^2}
\newcommand {\Dzero} {D\O\ }

\newcommand {\ppbar} {p\bar{p} }

\begin{document}

\title{ \boldmath Measurement of the $\lambdab$ Lifetime in
$\lambdabjpsilm$ in $\ppbar$ Collisions at $\sqrt{s}=1.96~{\rm TeV}$}

\affiliation{Institute of Physics, Academia Sinica, Taipei, Taiwan 11529, Republic of China} 
\affiliation{Argonne National Laboratory, Argonne, Illinois 60439} 
\affiliation{Institut de Fisica d'Altes Energies, Universitat Autonoma de Barcelona, E-08193, Bellaterra (Barcelona), Spain} 
\affiliation{Baylor University, Waco, Texas  76798} 
\affiliation{Istituto Nazionale di Fisica Nucleare, University of Bologna, I-40127 Bologna, Italy} 
\affiliation{Brandeis University, Waltham, Massachusetts 02254} 
\affiliation{University of California, Davis, Davis, California  95616} 
\affiliation{University of California, Los Angeles, Los Angeles, California  90024} 
\affiliation{University of California, San Diego, La Jolla, California  92093} 
\affiliation{University of California, Santa Barbara, Santa Barbara, California 93106} 
\affiliation{Instituto de Fisica de Cantabria, CSIC-University of Cantabria, 39005 Santander, Spain} 
\affiliation{Carnegie Mellon University, Pittsburgh, PA  15213} 
\affiliation{Enrico Fermi Institute, University of Chicago, Chicago, Illinois 60637} 
\affiliation{Joint Institute for Nuclear Research, RU-141980 Dubna, Russia} 
\affiliation{Duke University, Durham, North Carolina  27708} 
\affiliation{Fermi National Accelerator Laboratory, Batavia, Illinois 60510} 
\affiliation{University of Florida, Gainesville, Florida  32611} 
\affiliation{Laboratori Nazionali di Frascati, Istituto Nazionale di Fisica Nucleare, I-00044 Frascati, Italy} 
\affiliation{University of Geneva, CH-1211 Geneva 4, Switzerland} 
\affiliation{Glasgow University, Glasgow G12 8QQ, United Kingdom} 
\affiliation{Harvard University, Cambridge, Massachusetts 02138} 
\affiliation{Division of High Energy Physics, Department of Physics, University of Helsinki and Helsinki Institute of Physics, FIN-00014, Helsinki, Finland} 
\affiliation{University of Illinois, Urbana, Illinois 61801} 
\affiliation{The Johns Hopkins University, Baltimore, Maryland 21218} 
\affiliation{Institut f\"{u}r Experimentelle Kernphysik, Universit\"{a}t Karlsruhe, 76128 Karlsruhe, Germany} 
\affiliation{High Energy Accelerator Research Organization (KEK), Tsukuba, Ibaraki 305, Japan} 
\affiliation{Center for High Energy Physics: Kyungpook National University, Taegu 702-701, Korea; Seoul National University, Seoul 151-742, Korea; and SungKyunKwan University, Suwon 440-746, Korea} 
\affiliation{Ernest Orlando Lawrence Berkeley National Laboratory, Berkeley, California 94720} 
\affiliation{University of Liverpool, Liverpool L69 7ZE, United Kingdom} 
\affiliation{University College London, London WC1E 6BT, United Kingdom} 
\affiliation{Centro de Investigaciones Energeticas Medioambientales y Tecnologicas, E-28040 Madrid, Spain} 
\affiliation{Massachusetts Institute of Technology, Cambridge, Massachusetts  02139} 
\affiliation{Institute of Particle Physics: McGill University, Montr\'{e}al, Canada H3A~2T8; and University of Toronto, Toronto, Canada M5S~1A7} 
\affiliation{University of Michigan, Ann Arbor, Michigan 48109} 
\affiliation{Michigan State University, East Lansing, Michigan  48824} 
\affiliation{Institution for Theoretical and Experimental Physics, ITEP, Moscow 117259, Russia} 
\affiliation{University of New Mexico, Albuquerque, New Mexico 87131} 
\affiliation{Northwestern University, Evanston, Illinois  60208} 
\affiliation{The Ohio State University, Columbus, Ohio  43210} 
\affiliation{Okayama University, Okayama 700-8530, Japan} 
\affiliation{Osaka City University, Osaka 588, Japan} 
\affiliation{University of Oxford, Oxford OX1 3RH, United Kingdom} 
\affiliation{University of Padova, Istituto Nazionale di Fisica Nucleare, Sezione di Padova-Trento, I-35131 Padova, Italy} 
\affiliation{LPNHE, Universite Pierre et Marie Curie/IN2P3-CNRS, UMR7585, Paris, F-75252 France} 
\affiliation{University of Pennsylvania, Philadelphia, Pennsylvania 19104} 
\affiliation{Istituto Nazionale di Fisica Nucleare Pisa, Universities of Pisa, Siena and Scuola Normale Superiore, I-56127 Pisa, Italy} 
\affiliation{University of Pittsburgh, Pittsburgh, Pennsylvania 15260} 
\affiliation{Purdue University, West Lafayette, Indiana 47907} 
\affiliation{University of Rochester, Rochester, New York 14627} 
\affiliation{The Rockefeller University, New York, New York 10021} 
\affiliation{Istituto Nazionale di Fisica Nucleare, Sezione di Roma 1, University of Rome ``La Sapienza," I-00185 Roma, Italy} 
\affiliation{Rutgers University, Piscataway, New Jersey 08855} 
\affiliation{Texas A\&M University, College Station, Texas 77843} 
\affiliation{Istituto Nazionale di Fisica Nucleare, University of Trieste/\ Udine, Italy} 
\affiliation{University of Tsukuba, Tsukuba, Ibaraki 305, Japan} 
\affiliation{Tufts University, Medford, Massachusetts 02155} 
\affiliation{Waseda University, Tokyo 169, Japan} 
\affiliation{Wayne State University, Detroit, Michigan  48201} 
\affiliation{University of Wisconsin, Madison, Wisconsin 53706} 
\affiliation{Yale University, New Haven, Connecticut 06520} 
\author{A.~Abulencia}
\affiliation{University of Illinois, Urbana, Illinois 61801}
\author{J.~Adelman}
\affiliation{Enrico Fermi Institute, University of Chicago, Chicago, Illinois 60637}
\author{T.~Affolder}
\affiliation{University of California, Santa Barbara, Santa Barbara, California 93106}
\author{T.~Akimoto}
\affiliation{University of Tsukuba, Tsukuba, Ibaraki 305, Japan}
\author{M.G.~Albrow}
\affiliation{Fermi National Accelerator Laboratory, Batavia, Illinois 60510}
\author{D.~Ambrose}
\affiliation{Fermi National Accelerator Laboratory, Batavia, Illinois 60510}
\author{S.~Amerio}
\affiliation{University of Padova, Istituto Nazionale di Fisica Nucleare, Sezione di Padova-Trento, I-35131 Padova, Italy}
\author{D.~Amidei}
\affiliation{University of Michigan, Ann Arbor, Michigan 48109}
\author{A.~Anastassov}
\affiliation{Rutgers University, Piscataway, New Jersey 08855}
\author{K.~Anikeev}
\affiliation{Fermi National Accelerator Laboratory, Batavia, Illinois 60510}
% ...
\collaboration{CDF Collaboration}
\noaffiliation


\begin{abstract} 

We report a measurement of the $\lambdab$ lifetime in the exclusive
decay $\lambdabjpsilm$ in $p \overline{p}$ collisions at $\sqrt{s} = 
1.96~{\rm TeV}$ using an integrated luminosity of 1.0 ${\rm fb}^{-1}$ 
of data collected by the CDF II detector at the Fermilab
Tevatron. Using fully reconstructed  decays, we measure
$\tau(\lambdab) = 1.593 ~^{+0.083}_{-0.078} ~({\rm stat.}) \pm
0.033~{\rm (syst.)~ps}$. This is the single most precise measurement
of $\tau(\lambdab)$ and is 3.2 $\sigma$ higher than the current world
average.
\end{abstract}


\maketitle

\normalsize

The weak decay of quarks depends on fundamental parameters of the
standard model, including elements of the Cabibbo-Kobayashi-Maskawa
(CKM) matrix which describe mixing between quark families. Extraction
of these parameters from weak decays is complicated since observed
quarks are not free but are confined within color-singlet hadrons, as
described by quantum chromodynamics (QCD). An essential tool used in
this extraction is the heavy quark expansion (HQE) technique 
\cite{Manohar:2000dt}. In HQE, the total decay width of a heavy hadron
is expressed as an expansion in inverse powers of the heavy quark mass
$m_q$. Lifetime ratios of $b$-flavored hadrons are predicted to be
unity through $O(1/m_b)$, and $O(1/m^{2}_{b})$ corrections are small
($\simle$ 2\%) \cite{Bigi:1992su}. Detailed analysis of
$O(1/m^{3}_{b})$ corrections to the lifetime ratio lead to an expected
value of ${\tau (\lambdab)\over \tau (\bzero)}\simeq 0.94$
\cite{Uraltsev:1996ta}. This theoretical prediction has been in poor
agreement with measurements for more than a decade
\cite{Barate:1997if,Abe:1996df,Abazov:2004bn,Abreu:1999hu,Ackerstaff:1997qi}.
The current world average of the $\lambdab$ lifetime is $1.230 \pm
0.074 ~{\rm ps}$, corresponding to a ratio of ${\tau (\lambdab)\over
\tau (\bzero)} = 0.804 \pm 0.049 $~\cite{PDG}. Recent calculations
including next-to-leading-order QCD and $O(1/m^{4}_{b})$ corrections
lower the prediction \cite{Gabbiani:2004tp,Franco:2002fc}, making it
more consistent with the measurements.

In this Letter, we present a measurement of the $\lambdab$ lifetime,
$\tau({\lambdab})$, in the fully reconstructed decay $\lambdabjpsilm$,
with $\jpsimumu$ and $\lmppi$. Charge conjugate modes are implied
throughout. Our data sample consists of 1.0 fb$^{-1}$ of $p \overline
p$ collisions at $\sqrt{s} = 1.96 ~\rm TeV$ collected by the CDF II
detector at the Fermilab Tevatron between February 2002 and February
2006. This is the first measurement of $\tau({\lambdab})$ using fully
reconstructed decays that is competitive with the best previous
measurements, which are based on semileptonic decays. As compared to
fully reconstructed decays, measurements using partially reconstructed
semileptonic decays have additional uncertainties due to the missing
energy of the unobserved neutrino and the modeling of background from
other $b$-flavored baryons.

As a check of our method, we also measure $\tau({\bzero})$ using a
sample of $\bzerojpsiks$ decays, which has larger yield than the
$\lambdab$ sample. This decay channel is topologically similar to
$\lambdabjpsilm$, since both $\kshort$ and $\lam$ decay with large
displacement from the $b$-hadron-decay vertex. The analysis procedure
for $\lambdab$ is developed using $\bzerojpsiks$ only and checked
using other $b$-meson decays containing a $\jpsimumu$ in the final
state. The $\lambdab$ lifetime was not extracted until all procedures
were established including the estimate of the systematic uncertainty.

The components of the CDF II detector relevant to this analysis are
described briefly here; a more complete description can be found 
elsewhere \cite{Blair:1996kx}. Charged particles are reconstructed
using an open-cell drift chamber called the central outer tracker
(COT) \cite{COT} and 7 layers of silicon microstrip detectors with
radii between 2.4 cm and 28 cm \cite{SVXII}. These are immersed in a
1.4~T solenoidal magnetic field and cover the range $|\eta| \leq 1$,
where $\eta$ is the pseudorapidity defined as $\eta \equiv - \ln \tan
(\theta/2)$ and $\theta$ is the polar angle \cite{Tracking}. Four
layers of planar drift chambers (CMU) \cite{CMU} detect muons with
transverse momentum $\pt > 1.4 \GeVc$ within $|\eta|<0.6$. Additional
chambers and scintillators (CMX) \cite{CMX} cover $0.6 < |\eta| < 1.0$
for muons with $\pt > 2.0 \GeVc$.

A sample of $\jpsimumu$ candidates, collected using a dimuon trigger,
is selected to begin the reconstruction of $\lambdab$ and $\bzero$
candidates. At level 1 of a three-level trigger system, the eXtremely
Fast Tracker (XFT) \cite{XFT} uses COT information to fit
tracks. Those tracks with $\pt > 1.5 (2.0) \GeVc$ are extrapolated to
the CMU (CMX) chambers and compared with the positions of muon-chamber
tracks. The event passes level 1 if two or more XFT tracks are matched
to muon-chamber tracks. Opposite-charge and opening-angle requirements
are imposed at level 2. At level 3, full tracking information is used
to reconstruct $\jpsimumu$ candidates. Events with a candidate in the
mass range 2.7 to 4.0$\GeV$ are accepted at level 3 and permanently
recorded for further analysis.

Tracks from two triggered muons are constrained to originate from a
common vertex to make a $\jpsimumu$ candidate. To ensure a
high-quality vertex for the lifetime measurement, each muon track is 
required to have at least 3 axial hits in the silicon system. The
reconstructed $\mu^{+}\mu^{-}$ invariant mass is required to be in the
range $3.014 < M_{\muplus\muminus} < 3.174 \GeV$. This corresponds
to approximately $\pm 4 \sigma$ of the reconstructed width, which is
dominated by the resolution \cite{Acosta:2004yw}.

We construct $\kspipi$ and $\lmppi$ candidates from pairs of
oppositely-charged tracks fit to a common vertex. Since many  
$\kshort$ and $\lam$ decays occur outside some layers of the silicon
system due to their long lifetime, their tracks are not required to
have silicon hits. We suppress $\kshort$ and $\lam$ cross
contamination by rejecting $\kshort$ ($\lam$) candidates with
proton-pion (dipion) invariant mass in the range [1.1085,~1.1235]
([0.48175,~0.5115]) $\GeV$. 

The $\bzero$ and $\lambdab$ candidates are reconstructed by
associating $\jpsi$ candidates with $\kshort$ or $\lam$ candidates in
each event. We choose further selection requirements for our
$b$-hadron samples that optimize $S/\sqrt{S+B}$ where $S$ and $B$ are
the numbers of signal and background events, respectively. In the
optimization procedure, $S$ is estimated using a Monte Carlo
simulation, while $B$ is estimated using the $b$-hadron invariant mass
sidebands, which are chosen to exclude the data used in the lifetime
fits to avoid any potential bias.

\begin{figure}
\includegraphics[width=0.66\linewidth]{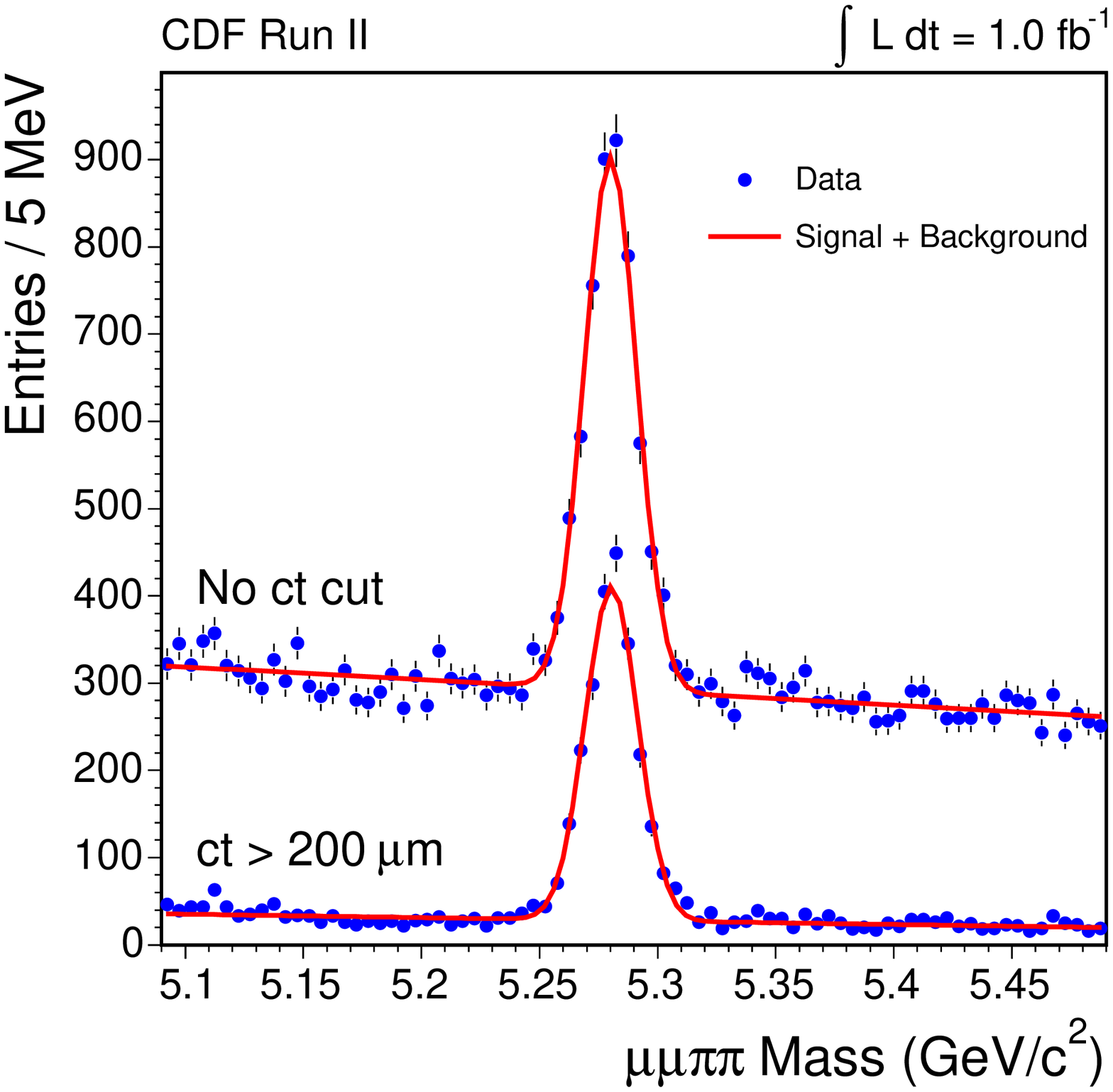}
\unitlength=0.96\linewidth
\put(-0.55,0.55){(a)} \\
\includegraphics[width=0.66\linewidth]{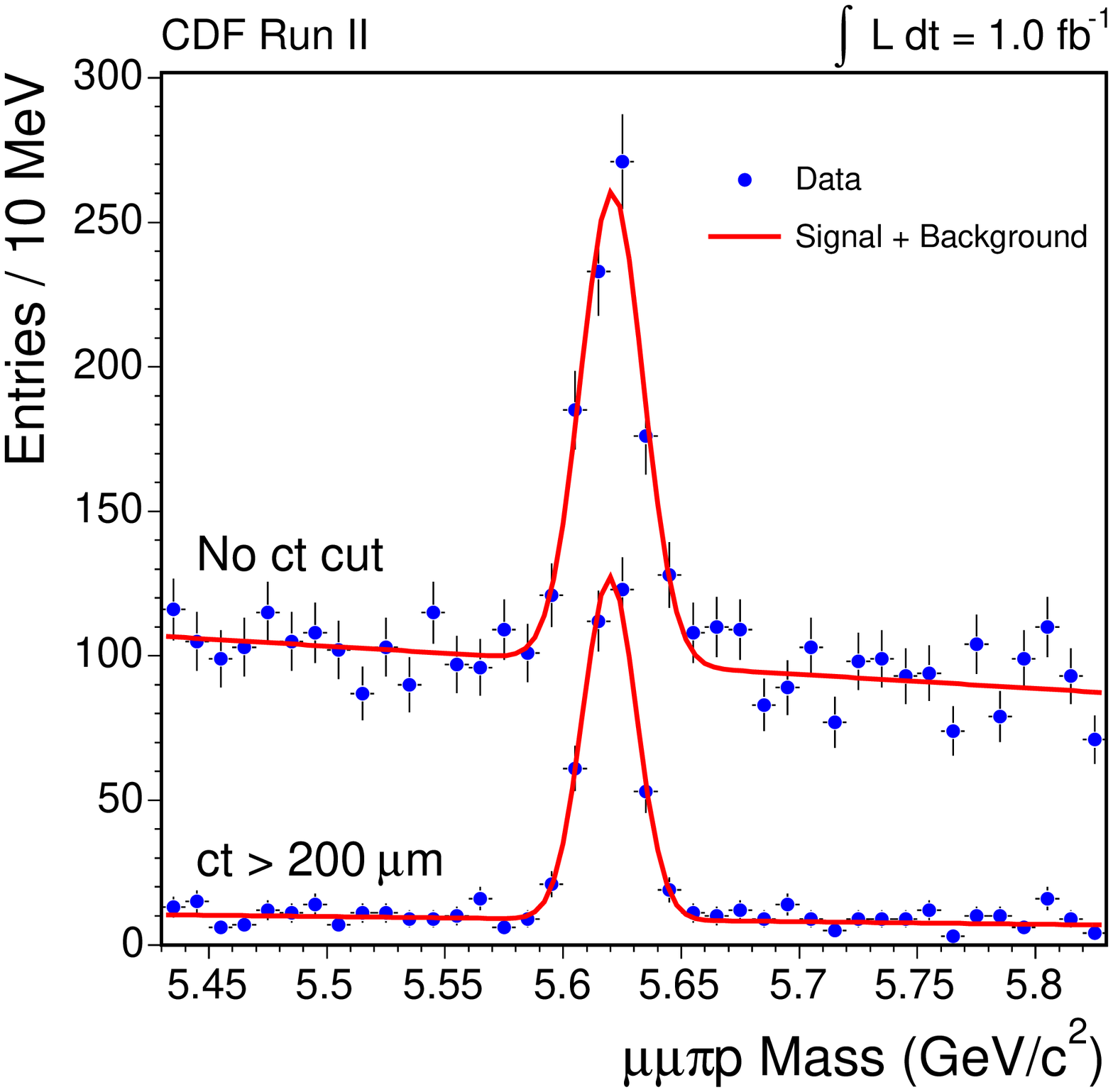}
\put(-0.55,0.55){(b)}
\caption{Invariant mass distribution of (a) $\bzerojpsiks$ and (b)
$\lambdabjpsilm$ candidates. The distributions with $ct > 200~\mum$,
where $t$ is the proper decay time, illustrate that the majority of
backgrounds originate from the primary interaction vertex. The
distributions are fit to the sum of a Gaussian signal and linear
background.}
\label{fig:fitMass}
\end{figure}

The selection requirements resulting from the optimization are 
described below. We require $0.473 < M_{\pi\pi} < 0.523~\GeV$ and
$\pt > 1.5\GeVc$ for $\kshort$ candidates. For $\lam$ candidates, we
require $1.107 < M_{p\pi} < 1.125~\GeV$ and $\pt > 2.6\GeVc$.
We require $\LxySigFromJpsi > 6$ for $\kshort$ and $\LxySigFromJpsi >
4$ for $\lam$, where $\LxyFromJpsi$ is defined as the 
distance from the $\jpsi$ vertex to the $\Vzero(\equiv \kshort$,
$\lam)$ vertex projected onto the $\Vzero$ transverse momentum vector
and $\LxyErrFromJpsi$ is its estimated uncertainty. Both $\bzero$ and
$\lambdab$ candidates are required to have $\pt > 4.0\GeVc$. Finally,
the $\chisq$ of a $b$-hadron kinematic fit is required to be less
than 26 for 5 degrees of freedom. This fit finds the best $b$-hadron
decay vertex and momentum subject to the constraints that the
muon tracks originate from a common vertex, the $\kshort$ or $\lam$
tracks originate from a common vertex with combined momentum pointing
back in three dimensions to the dimuon vertex, and the invariant
mass of the two muons is equal to the world average $\jpsi$ mass
\cite{PDG}. The invariant mass distributions of the $\bzero$ and
$\lambdab$ candidates passing these requirements are shown in
Fig.~\ref{fig:fitMass}. The yields are $N(\bzerojpsiks) = 3376 \pm 88$
(stat.) and $N(\lambdabjpsilm) = 538 \pm 38$ (stat.).

The lifetime of a $\bzero$ and $\lambdab$ is determined from the
distribution of proper decay time $t$ given by 
$\PDL \equiv {\LxyB}/{(\beta\gamma)_{\rm T}^b} = \LxyB {c~{\rm M}_b}/{\ptB},$
where $\LxyB$ is the distance traveled by each $b$-hadron candidate
along the direction of its transverse momentum $\ptB$ and
$(\beta\gamma)_{\rm T}^b \equiv \ptB / (cM_b)$ is the transverse boost,
where $M_b$ is the world average mass of the $b$ hadron \cite{PDG}. 
Since the $\jpsi$ vertex occurs at the same point as the $b$ hadron decay
and is well determined, it is used as the $b$-hadron decay vertex.
The $b$ hadron is assumed to originate from the average beamline
determined on a run-by-run basis using inclusive jet data.
The primary vertex for a given event is the $x-y$ position of this 
beamline at the average $z$ coordinate of the muon tracks at their 
closest approach to the beamline.

The lifetimes are extracted using the maximum likelihood method. The
likelihood function ${\cal L}$ is multivariate, and is constructed 
from the products of single variable probability density functions
describing the distributions of the invariant mass $m_i$, $ct_i$, and
their respective estimated resolutions $\sigma^{m}_i$ and $\sigma^{ct}_i$.
It is given by 
\begin{align*}
{\cal L} = \prod_{i=1}^{N} &\left[ (1-f_B)~{\cal
P}_S^{\PDL}(\ctauData | \ctauErrData)~{\cal
P}_S^{\sigma^{ct}}(\ctauErrData)~{\cal P}_S^m (\massData |
\massErrData) \right. \\
& \left. + ~f_B  ~{\cal P}_B^{\PDL}(\ctauData
| \ctauErrData)  ~{\cal P}_B^{\sigma^{ct}}(\ctauErrData)
 ~{\cal P}_B^m (\massData | \massErrData) \right],
\label{eq:pdf}
\end{align*}
where $N$ is the number of events in the $b$-hadron mass window, 
$f_B$ is the background fraction, and ${\cal P}^{\PDL}$, ${\cal
P}^{\sigma^{ct}}$, and ${\cal P}^m$ are probability density
functions for $\PDL$, $\sigma^{ct}$, and mass, respectively. The
mass resolution probability distributions 
${\cal P}^{\sigma^m}(\massErrData)$ do not appear in ${\cal L}$
because they are equal for signal and background, within the available
statistics. Since this 
is not true for the $\PDL$ resolution distributions, ${\cal
P}^{\sigma^{ct}}$ must be included in ${\cal L}$ \cite{punzi}. 

The mass distribution is modeled as the sum of a Gaussian signal and
linear background, where the Gaussian width $\massErrData$ is scaled
by a parameter to account for misestimation of the mass
resolutions. The $\PDL$ distribution is modeled by the sum of five
components, all convoluted with a Gaussian resolution function with a
scale factor parameter for the $\ctauErrData$: a positive exponential
($e^{-\ctauData/\ctau} / \ctau$) for the signal, a $\delta-$function
representing the zero-lifetime component, one negative and two
positive exponentials accounting for mismeasured decay vertices and
background from other heavy-flavor decays. The relative contribution
of each of the background components is determined by the fit. The
$\sigma^{ct}$ distribution is modeled by a Gaussian convoluted with an
exponential for both signal and background.

\begin{figure*}
\includegraphics[width=0.32\textwidth]{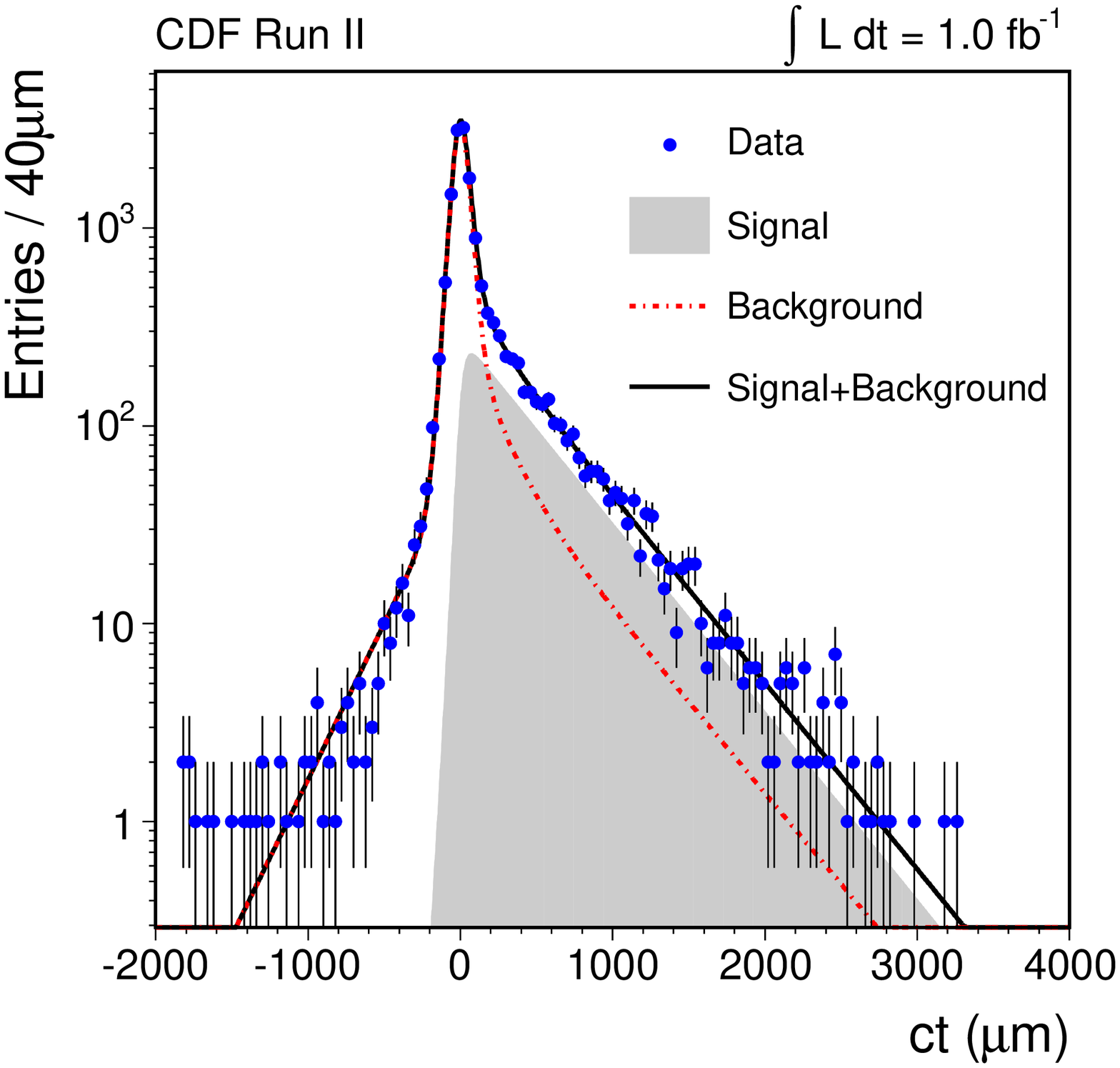}
\unitlength=0.32\textwidth
\put(-0.8,0.8){(a)}
\includegraphics[width=0.32\textwidth]{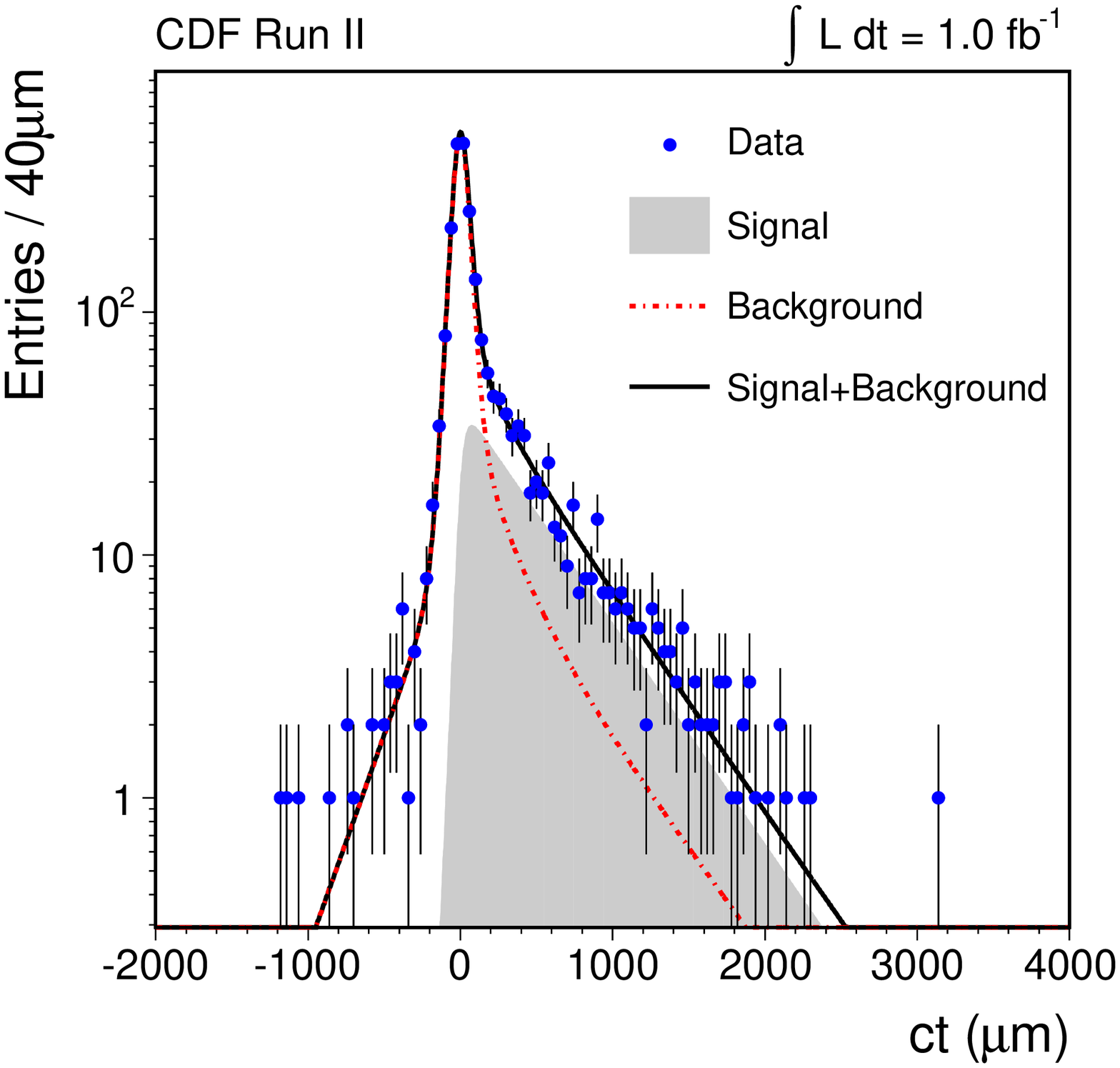}
\put(-0.8,0.8){(b)}
\includegraphics[width=0.32\textwidth]{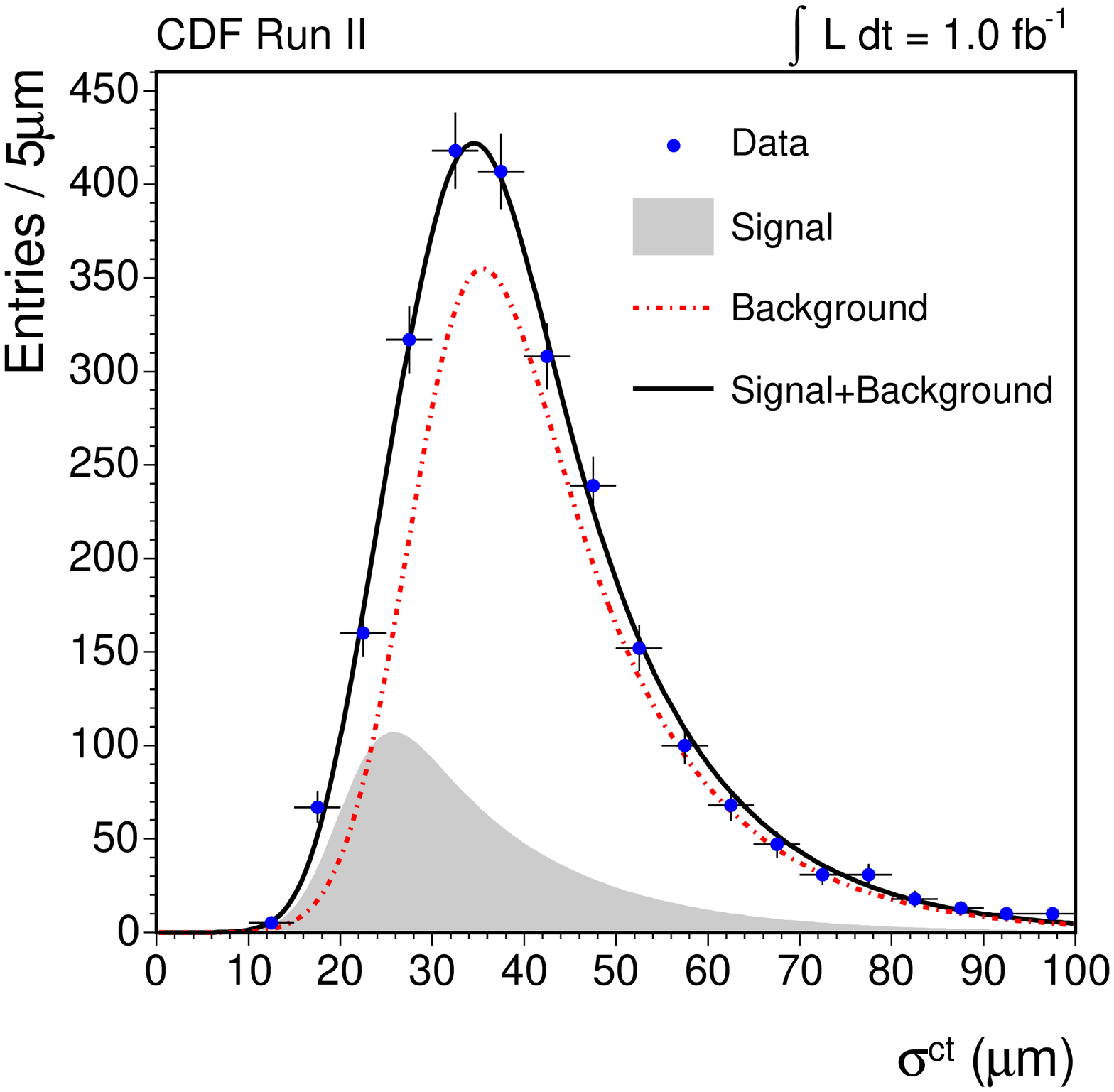}
\put(-0.8,0.8){(c)}
\caption{ (a) $\PDL$ fit projection for $\bzerojpsiks$ candidates;
and (b) $\PDL$ and (c) $\sigma^{\PDL}$  fit projections for
$\lambdabjpsilm$ candidates.}
\label{fig:fit}
\end{figure*}

We fit over the mass range [5.170,~5.390] and [5.521,~5.721]~$\GeV$
for $\bzero$ and $\lambdab$, respectively. These ranges provide a sufficient
sideband to constrain the background shape while avoiding regions
where the mass distribution has complex structure. For both $\bzero$
and $\lambdab$, we require $\massErrData < 20 \MeV$ and fit over the range
[-2000,~4000] $\mum$ in $\PDL$ and [0,~100] $\mum$ in $\sigma^{ct}$. We
maximize the likelihood to determine the best values of all fit model
parameters, including the signal lifetimes $\ctau(\bzero) = 456.8
~^{+9.0}_{-8.9}~\mum$ and $\ctau(\lambdab)=  477.6
~^{+25.0}_{-23.4}~\mum$. Fit projections are shown in Fig.~\ref{fig:fit}. 

Systematic uncertainties come from four main sources: fitting
procedure and model, primary vertex determination, alignment of
detector elements, and $\kshort$ or $\lam$ pointing requirement in the
$\bzero$ or $\lambdab$ kinematic fit. The fitting bias is determined
using a simple Monte Carlo simulation in which events are distributed
according to the fit model. The bias was found to be less than 
0.4 $\mum$ and 0.5 $\mum$
for $\bzero$ and $\lambdab$, respectively. The systematic
uncertainties due to $\PDL$ resolution and mass resolution are
estimated by including additional Gaussian components to their
respective parts of the model in separate fits to the data and
observing the deviations from the nominal result. We estimate the
contribution from our mass background model by fitting with a uniform
rather than linear background shape. The systematic uncertainty due to
the $\PDL$ background model is estimated by fitting with two or four
background exponentials convoluted with the resolution function and
fitting with two, three, or four background exponentials without
convolution. We study a possible mass dependence in the $\PDL$
background shape by separately fitting for $\bzero$ ($\lambdab$)
lifetime in the following low and high mass regions: [5.170,~5.3225]
([5.521,~5.651]) and [5.2375,~5.390] ([5.591,~5.721]) $\GeV$. The
observed shifts are consistent with the statistical differences of the
two samples for both modes. We use the average shift of 1.9 $\mum$ and
4.1 $\mum$ for $\bzero$ and $\lambdab$, respectively, as an estimate
of the systematic uncertainty due to a mass-dependent $\PDL$
background. We estimate the systematic uncertainty due to our
$\sigma^{ct}$ and $\sigma^{m}$ distribution models by the observed
shift between simple Monte Carlo events generated with the data
distributions but fit with our model compared with simple Monte Carlo
events both generated and fit with our model. Using the same simple
Monte Carlo technique, we estimate the systematic uncertainty due to a
correlation between the $\PDL$ and $\sigma^{ct}$ for the background
component by generating simple Monte Carlo data sets with the
correlation observed in the data and fitting with our baseline model
where this correlation is absent.

We estimate the systematic uncertainty due to our primary vertex
determination by comparing different choices of the $z$ coordinate used to
evaluate the run-averaged beamline. We estimate uncertainties due
to any residual misalignments of the silicon detector using Monte
Carlo samples generated with radial displacements of individual
sensors (internal alignment) and relative translation and rotation of
the silicon detector with respect to the COT (global alignment). 
We also study the resolution and bias on the $\Vzero$ 
pointing to the $\jpsi$ vertex in data. If these were strongly $\PDL$-dependent,
the kinematic fit quality requirement could bias the $b$-hadron lifetime.
We observe no lifetime bias and assign uncertainties
of 0.6 $\mum$ for $\bzero$ and 5.4 $\mum$ for $\lambdab$ based on the
statistical precision of our study.

The systematic uncertainties are summarized in
Table~\ref{tbl:syst}. We obtain total systematic uncertainties of  
4.9 $\mum$ for $\bzero$ and 9.9 $\mum$ for $\lambdab$ by adding the
individual uncertainties in quadrature.

\begin{table}
\caption{Systematic uncertainties (in $\mum$) for the measurement of
$\ctau(\bzero)$ and $\ctau(\lambdab)$. The total uncertainties are the
individual uncertainties added in quadrature.
\label{tbl:syst}}
\begin{ruledtabular}
\begin{tabular}{lcc}
Source            & $\ctau(\bzero)$ & $\ctau(\lambdab)$ \\ \hline
Fitter Bias                                   &  0.4    &  0.5 \\
Fit Model:                                    &         &      \\
~~$\PDL$ Resolution                           &  3.1    &  5.5 \\
~~Mass Signal                                 &  0.7    &  2.3 \\
~~Mass Background                             &  0.1    &  0.1 \\
~~$\PDL$ Background                           &  0.5    &  0.7 \\
~~$\sigma^{\PDL}$ Distribution Modeling       &  0.1    &  0.2 \\
~~$\sigma^m$ Distribution Modeling            &  0.6    &  0.2 \\
~~Mass-$\PDL$ Background Correlation          &  1.9    &  4.1 \\
~~$\PDL$-$\sigma^{\PDL}$ Background Correlation &  0.3    &  1.3 \\
Primary Vertex Determination                  &  0.2    &  0.3 \\
Alignment                                     &         &      \\ 
~~SVX Internal                                &  2.0    &  2.0 \\
~~SVX/COT Global                              &  2.2    &  3.2 \\
$\Vzero$ Pointing                             &  0.6    &  5.4 \\ \hline
Total                                         &  4.9    &  9.9 \\ 
\end{tabular}
\end{ruledtabular}
\end{table}

A number of cross checks on the analysis procedure are performed. We
measure the $\bplus$ and $\bzero$ lifetimes which are statistically 
consistent
with the world average values in the following decay modes: $\bzero
\rightarrow \psi^{(\prime)} \kshort$, $\bzero \rightarrow
\psi^{(\prime)} \kstarzero (\kstarzerokpi)$, $\bplus \rightarrow
\psi^{(\prime)} \kplus$, and $\bplus \rightarrow \psi \kstarplus
(\kstarpluskspi)$,  with $\psi^{(\prime)} \rightarrow \muplus
\muminus$ and $\psi^{\prime} \rightarrow \psi \piplus \piminus$. We
search for unexpected lifetime dependence on the $\Vzero$ and
$b$-hadron kinematics, data-taking period, number of tracks in the
event, and use of silicon hits on $\Vzero$ daughter and muon tracks;
no dependence is observed. Finally, we determine the lifetime using
two alternative techniques which give results consistent with our
baseline fit: a $\PDL$-only binned fit applied to sideband-subtracted
data and a fit to the mass distribution in $\PDL$ bins which is
insensitive to details of the $\PDL$ background shape. 

In summary, we measure 
\begin{equation*}
\tau(\lambdab) = 1.593 ~^{+0.083}_{-0.078} ~({\rm stat.}) \pm
0.033~{\rm (syst.)} ~{\rm ps}.
\end{equation*}
As a cross check, we also measure $\tau({B^0}) = 1.524 \pm
0.030 ~({\rm stat.}) \pm 0.016 ~({\rm syst.}) ~{\rm ps}$ which is
consistent with the world average $\tau({B^0}) = 1.530 \pm
0.009~{\rm ps}$. Our measurement of $\tau(\lambdab)$ is consistent
with the \Dzero result in the same channel \cite{Abazov:2004bn} at the
1.7$\sigma$ level and is the first measurement using a fully
reconstructed mode that reaches a precision comparable with the
previous best measurements based on semileptonic decays of the
$\lambdab$. It is also comparable in precision to the current world
average, but is 3.2 $\sigma$ higher \cite{PDG}. Forming a ratio with
the world average $\bzero$ lifetime, we determine 
\begin{equation*}
\frac{\tau (\lambdab)}{\tau (\bzero)} = 1.041 \pm 0.057~{\rm (stat.+syst.)}.
\end{equation*}
This ratio is consistent with the higher end of the theory predictions 
\cite{Uraltsev:1996ta,Gabbiani:2004tp,Franco:2002fc}.

We thank the Fermilab staff and the technical staffs of the
participating institutions for their vital contributions. This work
was supported by the U.S. Department of Energy and National Science
Foundation; the Italian Istituto Nazionale di Fisica Nucleare; the
Ministry of Education, Culture, Sports, Science and Technology of
Japan; the Natural Sciences and Engineering Research Council of
Canada; the National Science Council of the Republic of China; the
Swiss National Science Foundation; the A.P. Sloan Foundation; the
Bundesministerium f\"ur Bildung und Forschung, Germany; the Korean
Science and Engineering Foundation and the Korean Research Foundation;
the Particle Physics and Astronomy Research Council and the Royal
Society, UK; the Institut National de Physique Nucleaire et Physique
des Particules/CNRS; the Russian Foundation for Basic Research; the
Comisi\'on Interministerial de Ciencia y Tecnolog\'{\i}a, Spain; the
European Community's Human Potential Programme under contract
HPRN-CT-2002-00292; and the Academy of Finland.

\bibliography{document}

\begin{thebibliography}{20}
\bibitem[{Tra()}]{Tracking}
\bibinfo{note}{CDF II uses a cylindrical coordinate system in which $\varphi$
  is the azimuthal angle, $r$ is the radius from the nominal beamline, and $z$
  points in the proton beam direction with the origin at the center of the
  detector. The transverse plane is the plane perpendicular to the $z$ axis.}
\end{thebibliography}
\end{document}